\documentclass[conference]{IEEEtran}
\IEEEoverridecommandlockouts

\usepackage{cite}
\usepackage{amsmath,amssymb,amsfonts}
\usepackage{algorithmic}
\usepackage{graphicx}
\usepackage{textcomp}
\usepackage{url}
\usepackage{hyperref}
\usepackage{subcaption}
\usepackage{orcidlink}
\usepackage{adjustbox}
\usepackage{float}
\usepackage[most]{tcolorbox}
\usepackage{enumitem}
\usepackage{multirow}
\usepackage{booktabs}
\usepackage{lipsum}
\usepackage{caption}
\usepackage[utf8]{inputenc}
\usepackage{newunicodechar}
\newunicodechar{−}{\ensuremath{-}}
\usepackage{cite}
\usepackage{subcaption}
\usepackage{colortbl}
\usepackage{multirow}
\usepackage{booktabs}
\usepackage{xcolor}
\usepackage{pgfplots}
\pgfplotsset{compat=1.18}
\usepackage[table,xcdraw]{xcolor}
\usepackage{multirow}
\usepackage{booktabs}
\def\BibTeX{{\rm B\kern-.05em{\sc i\kern-.025em b}\kern-.08em
    T\kern-.1667em\lower.7ex\hbox{E}\kern-.125emX}}

\usepackage[T1]{fontenc}

\def\BibTeX{{\rm B\kern-.05em{\sc i\kern-.025em b}\kern-.08em
    T\kern-.1667em\lower.7ex\hbox{E}\kern-.125emX}}
\begin{document}

\title{LLM-based Automated Architecture View Generation: Where Are We Now?\\
}
\author{\IEEEauthorblockN{Miryala Sathvika
\orcidlink{0009-0000-2010-5000}
} 
\IEEEauthorblockA{
\textit{Software Engineering Research Centre} \\
\textit{IIIT Hyderabad, India}\\
\textit{miryala.sathvika@research.iiit.ac.in}}
\and
\IEEEauthorblockN{Rudra Dhar
\orcidlink{0000-0001-5206-6042}
}
\IEEEauthorblockA{
\textit{Software Engineering Research Centre} \\
\textit{IIIT Hyderabad, India}\\
\textit{rudra.dhar@research.iiit.ac.in}}
\and
\IEEEauthorblockN{Karthik Vaidhyanathan 
\orcidlink{0000-0003-2317-6175}
}
\IEEEauthorblockA{
\textit{Software Engineering Research Centre} \\
\textit{IIIT Hyderabad, India}\\
\textit{karthik.vaidhyanathan@iiit.ac.in}}
 }
 
\newcommand{\magenta}[1]{\textcolor{magenta}{#1}}
\newcommand{\blue}[1]{\textcolor{blue}{#1}}
\newcommand{\Human}{\stackon{\faUser}{\textcolor{green}{\faCheck}}}
\newcommand{\NoHuman}{\stackon{\faUser}{\textcolor{red}{\faTimes}}}
\vspace{-14pt}
\maketitle

\begin{abstract}
\textit{Context:} Architecture views are essential for software architecture documentation, yet their manual creation is labor-intensive and often leads to outdated artifacts. As systems grow in complexity, the automated generation of views from source code becomes increasingly valuable.

\noindent \textit{Goal:} We empirically evaluate the ability of LLMs and agentic approaches to generate architecture views from source code.

\noindent \textit{Method:} We analyze 340 open-source repositories across 13 experimental configurations using 3 LLMs with 3 prompting techniques and 2 agentic approaches, yielding 4,137 generated views. 
We evaluate the generated views by comparing them with the ground-truth using a combination of automated 
metrics 
complemented by human evaluations.

\noindent \textit{Results:} Prompting strategies offer marginal improvements. Few-shot prompting reduces clarity failures by 9.2\% compared to zero-shot baselines. The custom agentic approach consistently outperforms the general purpose agent, achieving the best clarity (22.6\% failure rate) and level of detail success (50\%). 

\noindent \textit{Conclusions:} LLM and agentic approaches demonstrate capabilities in generating syntactically valid architecture views. However, they consistently exhibit granularity mismatches, operating at the code level rather than architectural abstractions. This suggests that there is still a need for human expertise, positioning LLMs and agents as assistive tools rather than autonomous architects.

\end{abstract}


\section{Introduction}

Software architecture serves as the crucial bridge between requirements and implementation~\cite{bass2021software}, with architecture views enabling the design and communication of complex systems through visual representations~\cite{krutchen}. As defined by Rozanski and Woods, \textit{"A view is a representation of one or more structural aspects of an architecture that illustrate how the architecture addresses one or more concerns held by one or more of its stakeholders"}~\cite{rozanski2005software}. Architecture views enable the separation of concerns, breaking down multidimensional structures into comprehensible representations addressing specific stakeholders needs~\cite{clements2011documenting,architecturalviewpoints}. Recent analysis of 15,000 architectural views from 12,200 open-source projects reveal persistent challenges: 75\% of views are never updated after creation\textcolor{blue}{~\cite{migliorini2024}}. Maintenance remains predominantly manual and labor-intensive. Architectural knowledge is often scattered across code comments and commit messages rather than being  maintained as dedicated artifacts~\cite{kalliamvakou2016github,architecturedegradation,malavolta2021architecture}. 


Owing to the recent advancements in Generative AI, Large Language Models (LLMs) have been used to automate and support various aspects of architectural activities ranging from decision support to automated code and runtime architectural adaptation~\cite{esposito2025generative}. However, as mentioned by Esposito et al., the capabilities of LLMs in generating architectural views have been largely unexplored\textcolor{blue}{~\cite{esposito2025generative}}.
While LLMs excel at generating code snippets and completing functions, architecture views demand fundamentally different capabilities. This includes understanding system-wide concerns, identifying appropriate abstraction levels, recognizing architectural patterns, and addressing stakeholder-specific needs~\cite{iso42010_2022,schmid}. General-purpose coding agents capable of autonomous task decomposition and iterative tool use with memory capabilities (eg. Codex, etc) have recently been used to address complex multi-step challenges. Yet, the extent to which these LLMs and agents can effectively bridge the gap from implementation details to architectural view remains largely unexplored.


In this context, the goal of the paper is to perform an empirical study to investigate the capabilities and limitations of LLM-based approaches for generating architecture views directly from source code. These approaches include LLM prompting, a general-purpose agent, and a custom agentic approach, ArchView, which integrates architectural domain knowledge, concern specifications, and ISO/IEC/IEEE 42010-aligned architecture view descriptions~\cite{iso}. We conducted experiments on \textbf{340 }repositories across \textbf{13} experimental configurations, which resulted in \textbf{4,137} generated views.

We employ a combination of automated and human evaluation to capture the structural and architectural information in the generated views. Automated evaluation uses LLM-as-a-Judge for capturing Clarity, Completeness, and Consistency ~\cite{llmbasedevals,opportunities,surveyllmasjudge}, while image similarity metrics (SSIM~\cite{SSIMcitation}, PSNR~\cite{PSNRcitation}, SAM~\cite{SAMcitation}) capture visual correspondence. We complement this with human evaluation\footnote{human evaluation samples: \url{https://sa4s-serc.github.io/view_evaluation/}} as suggested by Oliveira et al. ~\cite{llmbasedevals}, focusing on accuracy and level of detail.

Our results show that LLMs can generate syntactically valid architecture views, but substantial quality gaps remain. Prompting approaches yield incremental improvements with diminishing returns. Clarity-related failures decrease from 40.0\% in zero-shot to 31.0\% in one-shot and 30.8\% in few-shot. Claude Code, a general-purpose coding agent, yielded the least effective results despite its tool access. ArchView achieves the best overall performance. 
The main contributions of this study are as follows.
\begin{itemize}
    \item A \textbf{large-scale empirical evaluation} of LLM-based architecture view generation across \textbf{340 open-source repositories} and\textbf{ 13 experimental configurations}, yielding a dataset of \textbf{4,137 generated views}.
    \item An analysis of model performance across specific architectural concerns and quality attributes presenting observations based on experimental results for automated view generation using LLM-based approaches.
    \item A \textbf{comprehensive replication package}, including all datasets, prompts, and scripts to ensure reproducibility and support future    research.
\end{itemize}
The remainder of this paper is organized as follows. Section~\ref{sec:background} provides background context, while Section~\ref{sec:studydesign} details our study design. Our empirical results and subsequent discussion are presented in Sections~\ref{sec:results} and \ref{sec:discussions} respectively. We address threats to validity in Section~\ref{sec:threatstovalidity}, discuss related work in Section~\ref{sec:related}, and conclude in Section~\ref{sec:conclusionandfutureworks}
\footnote{The replication package is available at: \url{https://zenodo.org/records/18772573}}.

\section{Background}
\label{sec:background}
\subsection{Prompting Techniques}

Zero-shot and few-shot prompting approaches require no additional training unlike fine-tuning, which involves retraining a language model on task-specific data~\cite{finetuning}. Zero-shot prompting provides the LLM with task instructions only~\cite{zeroshot}, whereas few-shot prompting (also known as in-context learning) incorporates several examples within the prompt to guide the model's behavior~\cite{fewshot}.
One-shot prompting is a special case of few-shot prompting where only one example is given to the LLM.
We use these different prompting strategies to perform our study along with other candidates.
\subsection{LLMs and Agents}
LLMs based on transformer architecture are trained on vast amounts of text data to understand and generate human-like text. They have demonstrated remarkable capabilities in software engineering tasks, including code generation, documentation, and analysis. Agents augment LLMs with the ability to use external tools, have memory, break down complex tasks into manageable steps, and iteratively refine their outputs. Agents can be general purpose, which can assists with various tasks, or be specialized for specific domains~\cite{wang2025agents}. Specialized agents combine domain knowledge with targeted workflows to achieve better performance.


\section{Study Design}
\label{sec:studydesign}
\subsection{Goal}
In this study, we aim to evaluate the ability of LLM-based approaches to generate architecture views from source code. Using the Goal-Question-Metric approach \cite{basili1994gqm}, we formalize our goal to:
\textbf{Analyze} the effectiveness of LLMs and agents \textbf{for the purpose of} generating software architecture views from 
source code, \textbf{from the viewpoint of} software architects, \textbf{in the context of} architectural knowledge management.

\subsection{Research Questions}

\noindent \textbf{RQ1: How do different LLM-based approaches compare for architecture view generation?}

This research question establishes baseline performance and compares all approaches systematically. We evaluate three LLMs across three prompting approaches and two agent-based approaches. We analyze the performance of LLM prompting strategies and investigate whether specialized, custom-built agents outperform both prompting and general-purpose agents.

\noindent \textbf{RQ2: How do architectural notation and granularity affect view generation quality?}

This research question examines how architectural characteristics influence generation effectiveness. We study how notation complexity (from simple boxes to intricate diagrams) and granularity affect the output quality of LLMs and agents. This may reveal LLM capabilities and limitations: which architectural representations LLMs naturally handle well versus which require specialized approaches or remain challenging despite domain knowledge integration.

\noindent \textbf{RQ3: How does performance vary across architectural concerns and quality attributes?}

This research question evaluates the effectiveness of generating distinct types of views by analyzing performance across diverse architectural concerns and quality attributes. By observing where current models perform well and where they face difficulties, we aim to offer practical insights that help practitioners choose suitable modeling approaches.
\begin{figure*}[t]
    \centering
\includegraphics[width=1\linewidth]{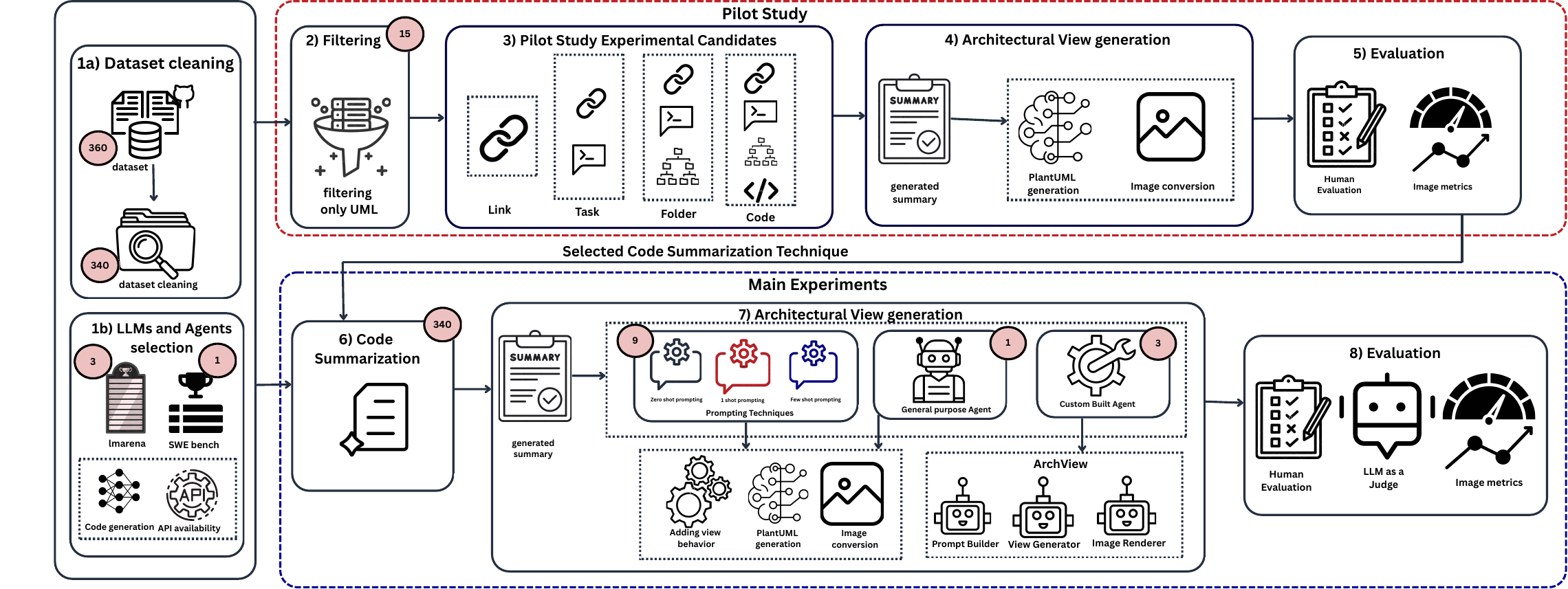}
        \caption{Study Diagram}
    \label{fig:study diagram}
\end{figure*}

\subsection{Pilot Study}
To generate an architectural view, it is necessary to extract information from the source code. Code summarization enables extraction of this information.
We conducted a pilot study to evaluate four code summarization techniques for extracting information~\cite{sourcecodemapping}, thereby identifying the most effective technique and the best zero-shot prompt template. The red box in fig.~\ref{fig:study diagram} highlights the pilot study.
\subsubsection*{1a) Dataset Cleaning}
We employed the dataset provided by Migliorini et al.~\cite{10.1007/978-3-031-70797-1_27}, which comprises of approximately 15,000 architecture views from 12,200 open-source GitHub repositories. From this dataset, we focused on the subset of 360 manually coded repositories with detailed annotations. 
\subsubsection*{1b) LLM Selection}
We selected LLMs based on the lmarena ~\footnote{\url{https://lmarena.ai/}} leaderboard~\cite{lmarena}. We use LLMs for two purposes: code summarization and view generation. For the pilot study, we employed the same LLM for both tasks. We selected the top 10 models from the leaderboard at the time of study (i.e., 07-06-2025), further filtered based on API availability and integration feasibility.  Table~\ref{tab:llm-comparison for pilot study} presents the selected models.
\subsubsection*{2) Filtering}
As shown in Fig.~\ref{fig:study diagram}, we applied filtering criteria to select views employing UML as the architectural notation, given that LLMs exhibit better comprehension of UML diagrams ~\cite{evaluatinguml}. Filtering resulted in 15 repositories.
\begin{table}[h]
\centering
\renewcommand{\arraystretch}{1.1}
\begin{footnotesize}
\begin{tabular}{|p{2.5cm}|p{2.1cm}|p{1.3cm}|p{1.3cm}|}
\hline
\textbf{Model} & \textbf{Context Window} & \textbf{Availability} & \textbf{License} \\
\hline
GPT-4o & 128K tokens & API & Proprietary \\
Gemini2.0-Flash & 1M tokens & API & Proprietary \\
\hline
\end{tabular}
\end{footnotesize}
\caption{Selected Models for Pilot Study}
\label{tab:llm-comparison for pilot study}
\end{table}

\subsubsection*{3) Pilot Study experimental candidates}
\begin{itemize}
    \item \textbf{Link:} This baseline technique provides only the GitHub repository URL to the LLM, relying on the model to access and process the repository independently. 
    \item \textbf{Task Specification:} Building on the baseline, this technique provides a detailed, structured prompt template with explicit instructions regarding the task objectives and expected I/O format.
    \item \textbf{Folder structure:} To improve on the above techniques, we provide the repository's directory structure in the context to give the LLM a high-level overview of the codebase. This organizational context may help the model infer relationships between files and reason about the system at a global level.
    \item \textbf{Code:} In this technique, the source code is provided along with the repository link and folder structure. When the code exceeds the context window limit, we employ a two-phase hierarchical summarization strategy: (1) the source code is partitioned into manageable chunks, (e.g.,  5,000 tokens for GPT), each summarized individually; (2) the chunk-level summaries are aggregated and summarized again to eliminate redundancy, consolidate information, and achieve higher-level abstractions.
\end{itemize}
\subsubsection*{4) Architectural view generation}
The extracted repository summary is supplied as context to the LLM to generate the corresponding view specification in PlantUML\footnote{\url{https://plantuml.com}}. The prompt specifies the diagram type (component diagram or sequence diagram) based on the behavioral characteristics of the ground truth view for the corresponding repository. This prompt template is used for zero-shot prompting in the experiments. The generated PlantUML code is converted to visual diagram locally using the PlantUML package. The evaluation methodology is detailed in Section~\ref{evalautionmetrics}. The prompts can be found in our \href{https://anonymous.4open.science/r/Viewgeneration-B2BF/pilot%20study/Code/chatgpt/code/View_generation.py}{replication package}.
\subsection{Key Findings from the Pilot Study}
We evaluated and compared the generated summaries and views across all four techniques and both LLMs. The pilot study yielded several critical insights that informed our main experimental design. First, LLMs occasionally produced syntactically incorrect PlantUML code, so a retry mechanism was introduced that supplies the erroneous output and corresponding compiler errors back to the model for up to three iterative corrections. Second, excessively large inputs increased failure rates with incoherent or incomplete views, motivating our hierarchical summarization approach to manage context window constraints. Finally, the generated views frequently lacked clear architectural concerns and showed poor alignment with the ground truth, reflected by a low SSIM score of 0.358 and consistent with our manual observations. This motivated the explicit incorporation of concern and behavior specifications into the generation prompts.
\subsection{Experiment Workflow}
\label{experimental flow}
\subsubsection*{1a) Dataset Cleaning}
We utilized the dataset provided by Migliorini et al.~\cite{migliorini2024}. From this dataset, we focused on the subset of 360 manually coded repositories with detailed annotations. Each view is manually coded according to multiple dimensions: (1) behavioral characteristics (static, dynamic, or both), (2) architectural notation (e.g., boxes-and-arrows, UML, icons), (3) architectural concerns addressed (e.g., deployment, security, control flow), (4) quality attributes according to ISO/IEC 25010 (e.g., maintainability, security, performance efficiency), (5) granularity level (high, medium, low), and (6) additional metadata including textual descriptions, component and connector characteristics, and technologies used.

Filtering ensured the inclusion of only repositories with accessible source code; inactive, missing, or archived links were excluded. The final dataset comprised 340 repositories.

The filtered dataset exhibits diverse characteristics: 54\% of views employ informal boxes-and-arrows notation, 49\% address static behavior while 41\% address dynamic behavior, and concerns span from general documentation (33\%) to deployment (14\%), control flow (26\%), and security (2\%). Quality attributes are dominated by maintainability (47\%) and functional suitability (14\%), with granularity distributed across high (58\%), medium (41\%), and low (10\%) levels. 
\subsubsection*{1b) LLMs and Agents Selection}
We require two LLMs for view generation. For summarization, we used Gemini-2.0-Flash, as it demonstrated the best performance in the pilot study. For view generation, we selected LLMs from the lmarena leaderboard based on their code completion capabilities. The study was conducted from April to September. Selection criteria were based on the top 10 models ranked during that period. From this set, two proprietary and one open-source model were selected to ensure diversity. Additionally, Claude Code agent was included for agent-based view generation due to its consistently high performance on SWE-bench~\footnote{\url{https://www.swebench.com}}, as well as its agentic capabilities for iterative code reasoning~\cite{Jimenez2023SWEbenchCL}. Table~\ref{tab:llm-comparison} presents the selected experimental candidates.
\begin{table}[h]
\centering
\renewcommand{\arraystretch}{1.1}
\begin{footnotesize}
\begin{tabular}{|l|p{2.9cm}|c|c|}
\hline
\textbf{Type} & \textbf{Experiment Candidate} & \textbf{Context Length} & \textbf{Open Source} \\
\hline
\multirow{3}{*}{LLM} & Deepseek V2.5 & 64K & Yes \\
 & Claude 3.5 Sonnet & 200K & No \\
 & GPT-4o & 128K & No \\
\hline
\multirow{2}{*}{Agent} & Claude Code(3.5 Sonnet) & API/CLI & No \\
 & ArchView & 64K & Yes \\
\hline
\end{tabular}
\end{footnotesize}
\caption{Selected experimental candidate}
\label{tab:llm-comparison}
\end{table}
\vspace{-4mm}
\subsubsection*{6) Code Summarization}
We chose code-based hierarchical summarization to generate repository summaries, as this proved most effective in the pilot study. We incorporated architectural concern specifications to provide behavioral context and improve view consistency.
\subsubsection*{7) Architectural View generation}
As shown in Fig.~\ref{fig:study diagram}, we evaluated five approaches (zero-shot, one-shot, few-shot, a general-purpose agent, and custom-built agent) resulting in 13 experimental configurations: three each for zero-shot, one-shot, and few-shot prompting (nine LLM-based settings), and four agentic approaches. Each approach receives the repository summary and the architectural concern specification as input. The generated view specification code is then compiled into visual diagrams and stored for evaluation. Complete prompts are provided in our \href{https://zenodo.org/records/18772573}{replication package}.

    \textbf{Zero-Shot Prompting (ZS):} This baseline approach provides summary extracted from source code and architectural concern specifications, along with task instructions.
    
    \textbf{One-Shot Prompting (OS):}As demonstrated above,  we augment the zero-shot template with a single example matching the target view's behavioral type (static or dynamic). Each example includes architectural information, concern specification, and corresponding PlantUML code.
    
    \textbf{Few-Shot Prompting (FS):} We provide 10 examples representing diverse view types and concerns additional to the task instructions, each with corresponding architectural information and PlantUML code.\\ 
    \textbf{General-Purpose Agent (GPA):} We evaluate a general-purpose coding agent, using the zero-shot prompt template to assess whether general coding capabilities suffice for architecture view generation.
    
\begin{tcolorbox}[colback=blue!10!white, 
    colframe=blue!80!black, 
    title=Prompt Template for View Generation,
    size=tight,        
    boxsep=1.5pt,      
    left=2pt, right=2pt, top=2pt, bottom=2pt, 
    arc=1.5mm,
    auto outer arc]
\label{promptemplate}
\textbf{Role Definition:} You are a software architect expert.\\
\textbf{Task Definition:} \\
You are given the summary of the code repository:
\textbf{\textit{\{Summary\}}}\\
and the architectural concern and behavior of the view:
\textbf{\textit{\{Concern,behavior\}}}.
Provide the PlantUML code corresponding to the summary addressing.\\
\rule{\linewidth}{0.5pt}
\noindent\textbf{One-shot:} \\
\textit{\{Example view's summary, concern, behavior\}}\\
\textit{\{Example view's PlantUML code\}}\\
\hrulefill
\rule{\linewidth}{0.5pt}
\noindent\textbf{Few-shot:} \\
\textit{\{Example1 view's summary, concern, behavior\}}\\
\textit{\{Example1 view's PlantUML code\}}\\
...\\
\textit{\{Example 10 view's summary, concern, behavior\}}\\
\textit{\{Example 10 view's PlantUML code\}}\\
\textit{If the generated code fails to compile, \textbf{\{Code\}} and \textbf{\{Error\}} are provided for correction.}
\textit{More detailed task description and output instructions are provided.}
\end{tcolorbox}

     \textbf{Custom-built Agent (ArchView (AV)):} Prior approaches for view generation often on hardcoded PlantUML as their output format. However, this is often far from the diverse, semi-formal notations used in the dataset. To address this gap, we developed ArchView, a multi-agent framework designed for dynamic generation. By moving beyond fixed templates, ArchView adapts its notation and structure based on the specific context. It employs a specialized toolset (\textit{\textbf{repository cloner, folder extractor, and code summarizer}}) to ensure high-fidelity, context-aware outputs. \\
    \textit{\textbf{Prompt Builder:}} It ensures the generated view is specifically tailored to the given source code. To construct the prompt, the agent uses (1) Architectural Concerns, and granularity; (2) Design Style Specifications; (3) Extracted Repository Metadata; and (4) View-Specific Instructions, tailored to the distinct type of view being generated.

\textbf{\textit{View Generator:}} It processes the structured prompt to produce the diagram code. The agent is notation-agnostic and can adapt its output based on the specific architectural requirements defined by the Prompt Builder.

\textbf{\textit{Image Renderer: }}This agent manages the compilation and validation of the generated code. To ensure robustness, it implements an error-feedback loop: if a syntax error occurs, the Renderer captures the compiler's feedback and re-tasks the View Generator to fix the code.

\subsubsection*{8) Evaluation Metrics}
\label{evalautionmetrics}

We employ a combination of automated and human evaluation approaches to comprehensively assess the architecture views. 
Following the framework proposed by Oliveira et al.~\cite{llmbasedevals}, we evaluate generated architecture views across five key dimensions: Clarity, Completeness, Consistency (3Cs), Accuracy, and Level of Detail. The authors demonstrated automated metrics using LLMs alone cannot adequately capture diagram granularity and accuracy, while human evaluation alone is resource-intensive and subject to individual bias. Therefore, we adopt a hybrid approach combining automated LLM-based assessment for the 3Cs with human evaluation for Accuracy and Level of Detail.

\textbf{LLM as a Judge:} We employ an LLM-as-a-Judge approach to evaluate the 3Cs ~\cite{zhengllm, gptscore, MLLMasajudge}:
\begin{itemize}

    \item \textit{Clarity} measures the comprehensibility and interpretability of the generated diagram for both technical and non-technical stakeholders, evaluating symbol and icon quality, label correctness, information presentation, layout logic, and component and connector specification.
    \item \textit{Completeness} measures whether the generated diagram includes all necessary architectural knowledge present in the ground truth without significant omissions or inappropriate additions, focusing on component and connector coverage penalizing extraneous and missing elements.
    \item \textit{Consistency} evaluates the uniform application of visual conventions, notational standards, and structural patterns throughout the diagram, as well as alignment with the ground truth reference, examining notational, styling, and naming consistency, structural and semantic alignment.
\end{itemize}


For each dimension, views are assigned one of three ratings: \textit{Meets Expectations} (no significant issues), \textit{Partially Meets Expectations} (minor issues requiring small improvements), or \textit{Does Not Meet Expectations} (major issues).

\textbf{Human Evaluation (Accuracy and Level of Detail):} We conducted human evaluation focusing two dimensions:
\begin{itemize}
    \item \textit{Accuracy} assesses whether the generated view correctly represents the architectural relationships and behaviors present in the ground truth.
    \item \textit{Level of Detail} evaluates whether the abstraction level is appropriate for the architectural view.
    
\end{itemize}

Based on automated LLM-as-a-Judge results (Table~\ref{tab:ratings_compact}), we selected the best-performing model from each of the five approaches (ZS, OS, FS, GPA, and AV) for human evaluation. Model selection prioritized those with the lowest percentage of "Does Not Meet Expectations" ratings aggregated across all three 3Cs dimensions.

The analysis comprises 110 architecture view evaluations, derived from the application of five approaches to a set of 22 statistically representative repositories. Samples were obtained through random sampling (85\% confidence level, $\pm$15\% margin of error), and two authors independently validated the output. We employed a blinded study design where evaluators received five generated views per sample (one from each experimental setting) alongside the corresponding ground truth, with no indication of which approach produced each view. For each sample, evaluators provided independent ratings using the three-level scale and documented  justifications. Results were averaged across both evaluators.

\textbf{Image-Based Metrics:} We quantitatively assess similarity between generated and ground truth diagrams using the \textit{Structural Similarity Index (SSIM)}. SSIM jointly evaluates luminance, contrast, and structural correspondence, where values approaching 1 indicate near-identical images. While we also computed \textit{Peak Signal-to-Noise Ratio (PSNR)}, S\textit{ignal to Reconstruction Error Ratio (SRE)}, and S\textit{pectral Angle Mapper (SAM)}, we report only SSIM here for brevity, the full suite of results is available in our replication package.\\
\textbf{Statistical Significance:} 
We performed statistical analysis to validate the observed differences in generation quality across the five approaches (We chose the best performing model in each). As our automated metrics rely on ordinal ratings, the assumption of normality was ruled out. For comparisons(RQ1), we utilized the Friedman test ($\alpha=0.05$)~\cite{friedman}, because our setup involved paired observations where all approaches were evaluated against the same set of 340 repositories. Upon finding statistical significance, we performed Wilcoxon signed-rank tests~\cite{wilcoxon} with Bonferroni correction then performed Matched-Pairs Rank-Biserial Correlation (r) as the effect size~\cite{kerby}, interpreted as small (0.1), medium (0.3), or large (0.5). Consequently (RQ2 and RQ3), we employed the Kruskal--Wallis~\cite{kruskhal} $H$-test ($\alpha = 0.05$) as our non-parametric omnibus test to identify significant differences across more than two independent groups. If there is a significant difference,  we proceeded to conduct a post-hoc Dunn’s test~\cite{Dunns} with Bonferroni correction (AV vs. GPA, AV vs. FS, AV vs. OS, and AV vs. ZS, $\alpha_{adj} = 0.0125$ ($0.05/4$)). To quantify the magnitude of these differences, we calculated Cliff’s Delta ($\delta$) effect sizes~\cite{cliff}.

\section{Results}
\label{sec:results}
\definecolor{medGreen}{HTML}{C6E0B4}  
\definecolor{medOrange}{HTML}{F8CBAD} 

\begin{table*}[ht]
\centering
\renewcommand{\arraystretch}{1.25} 
\scriptsize
\setlength{\tabcolsep}{4pt} 

\begin{tabular}{|l|l|l|c|c|c|c|c|c|c|c|c|c|c|c|}
\hline
\multirow{2}{*}{\textbf{Type}} & 
\multirow{2}{*}{\textbf{Approach}} & 
\multirow{2}{*}{\textbf{Candidate}} & 
\multirow{2}{*}{\textbf{Total}} & 
\multicolumn{3}{c|}{\textbf{Clarity (\%)}} & 
\multicolumn{3}{c|}{\textbf{Completeness (\%)}} & 
\multicolumn{3}{c|}{\textbf{Consistency (\%)}} & 
\multirow{2}{*}{\textbf{SSIM}} & 
\multirow{2}{*}{\textbf{Error}} \\ 

\cline{5-13}
 & & & & \textbf{M} & \textbf{P} & \textbf{N} 
 & \textbf{M} & \textbf{P} & \textbf{N} 
 & \textbf{M} & \textbf{P} & \textbf{N} 
 & & \\ \hline

\multirow{9}{*}{\textbf{LLM}} 
& \multirow{3}{*}{Zero-shot} 
  & DeepSeek-V2.5      & 320 & 0.3 & 50.4 & 49.3 & 0.3 & 17.0 & \cellcolor{medGreen}\textbf{82.7} & 0.3 & 10.4 & 89.3 & 0.56 & 20 \\
& & GPT-4o        & 324 & 0.6 & 52.5 & 46.9 & 0.3 & 11.4 & 88.3 & 0.3 & 11.7 & 88.0 & 0.58 & 16 \\
& & \textbf{Claude-3.5 Sonnet}  & 325 & 1.0 & 59.0 & \cellcolor{medGreen}\textbf{40.0} & 0.0 & 17.1 & 82.9 & 0.0 & 15.6 & \cellcolor{medGreen}\textbf{84.4} & 0.58 & 15 \\ \cline{2-15} 

& \multirow{3}{*}{One-shot} 
  & \textbf{DeepSeek-V2.5}      & 320 & 1.8 & 67.2 & \cellcolor{medGreen}\textbf{31.0} & 0.6 & 20.9 & 78.5 & 0.6 & 21.5 & \cellcolor{medGreen}\textbf{77.9} & 0.56 & 20 \\
& & GPT-4o       & 324 & 0.6 & 63.3 & 36.1 & 0.6 & 17.9 & 81.5 & 0.6 & 17.0 & 82.4 & 0.58 & 16 \\
& & Claude-3.5 Sonnet  & 310 & 1.3 & 57.1 & 41.6 & 0.6 & 18.4 & \cellcolor{medGreen}\textbf{81.0 }& 0.3 & 11.6 & 88.1 & 0.58 & 30 \\ \cline{2-15} 

& \multirow{3}{*}{Few-shot} 
  & DeepSeek-V2.5      & 308 & 1.5 & 64.1 & 34.4 & 0.9 & 21.4 & 77.7 & 0.6 & 18.7 & 80.7 & 0.55 & 32 \\
& & GPT-4o         & 312 & 2.2 & 64.4 & 33.3 & 1.6 & 17.6 & 80.8 & 0.6 & 22.8 & \cellcolor{medGreen}\textbf{76.6} & 0.57 & 28 \\
& & \textbf{Claude-3.5 Sonnet} & 299 & 1.0 & 68.2 & \cellcolor{medGreen}\textbf{30.8} & 0.0 & 26.4 & \cellcolor{medGreen}\textbf{73.6} & 0.0 & 17.7 & 82.3 & 0.59 & 41 \\ \hline

\multirow{4}{*}{\textbf{Agent}} 
& GPA
  & Claude Code        & 324 & 0.6 & 27.6 & \cellcolor{medOrange}\textbf{71.8} & 0.9 & 16.3 & \cellcolor{medOrange}\textbf{82.8} & 0.0 & 9.8 & \cellcolor{medOrange}\textbf{90.2} & 0.55 & 16 \\ \cline{2-15} 

& \multirow{3}{*}{ArchView}
  & DeepSeek-V2.5      & 322 & 1.2 & 59.3 & 39.4 & 0.3 & 11.5 & 88.2 & 0.3 & 21.4 & 78.3 & \cellcolor{medGreen}\textbf{0.64} & 18 \\
& & GPT-4o         & 324 & 1.5 & 65.6 & 32.8 & 0.0 & 14.6 & 85.4 & 0.0 & 18.9 & 81.1 & 0.62 & 16 \\
& & \textbf{Claude-3.5 Sonnet} & 325 & 2.5 & 74.8 & \cellcolor{medGreen}\textbf{22.6} & 0.0 & 15.0 & \cellcolor{medGreen}\textbf{85.0} & 0.0 & 27.4 & \cellcolor{medGreen}\textbf{72.6} & 0.60 & \cellcolor{medGreen}\textbf{15} \\ \hline
\end{tabular}

\caption{Percentage distribution of ratings for 3Cs, generation errors, and mean SSIM scores. \textbf{M} = Meets, \textbf{P} = Partially Meets, \textbf{N} = Does Not Meet. \colorbox{medGreen}{Green cells} denote best performance, while \colorbox{medOrange}{Orange cells} indicate significant failures.}
\label{tab:ratings_compact}
\end{table*}

\subsection*{RQ1: How do different LLM-based approaches compare for architecture view generation?}

\begin{figure}[!htbp]
\centering
\includegraphics[width=\columnwidth]{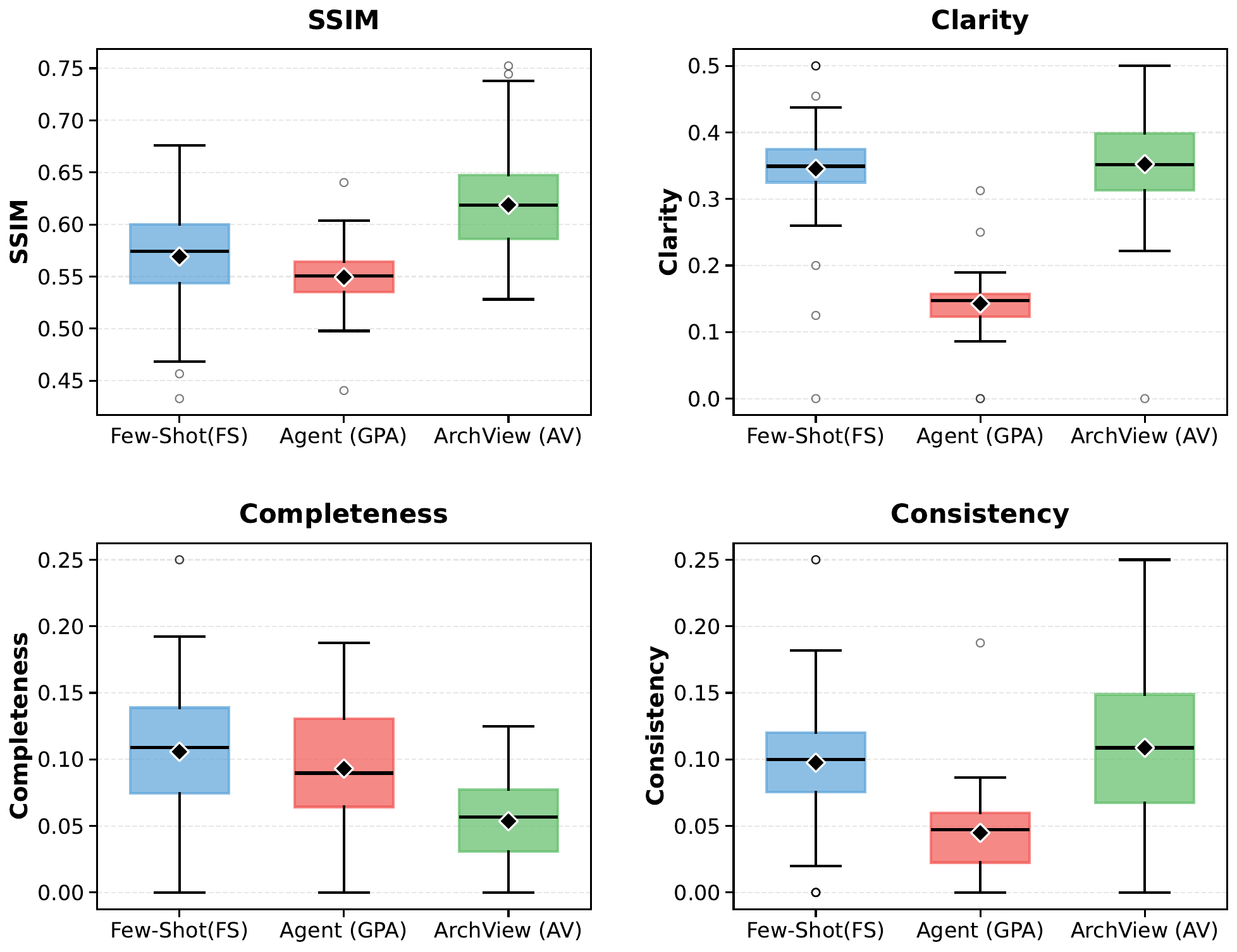}
\caption{Overall strategy comparison across SSIM, LLM Composite, Clarity, and Completeness metrics. }
\label{fig:overall_comparison}
\end{figure}

As shown in Table~\ref{tab:ratings_compact}, Claude-3.5 Sonnet achieved the highest performance within the Zero-Shot (ZS) configuration, exhibiting 40.0\% clarity failure rate. All the models performed poorly on completeness (>82\% failures) and consistency (>84\% failures). SSIM values remained moderate. Human evaluation reflected similar weaknesses (Table~\ref{tab:consensus_by_setting}).\\
OS prompting produced largest clarity gains with the best performing model improved by 18.3 points(49.3\% to 31.0\%). Completeness improved modestly, with DeepSeek-V2.5 at 78.5\% failures, ChatGPT-4o (81.5\%), and Claude-3.5 (81.0\%).\\
Using FS prompting approach, Claude-3.5 Sonnet achieved the best clarity and completeness, though improvement is low compared to the ZS to OS transition. Human evaluations diverged sharply: despite its best clarity, FS Claude-3.5 Sonnet achieved 45.0\% accuracy and 60.0\% level-of-detail success.\\
GPA (Claude code) produced worst performance across all 13 configurations (Table~\ref{tab:ratings_compact}). Clarity failures reached 71.8\%, 41.0 point worse than FS Claude-3.5 Sonnet. Consistency failures reached 90.2\%. Despite sophisticated tool-use and memory capabilities, completeness performance (82.8\% failures) merely matched ZS approach, and error rates matched ZS levels. SSIM at 0.55 was lowest among all approaches. Human evaluation confirmed these results with 0.0\% accuracy and level-of-detail success.
\begin{table}[ht]
\centering
\renewcommand{\arraystretch}{1.3}
\resizebox{\columnwidth}{!}{  
\begin{tabular}{|l|l|c|c|c|c|c|c|c|}
\hline
\multirow{2}{*}{\textbf{Approach}} & \multirow{2}{*}{\textbf{Candidate}} & \multirow{2}{*}{\textbf{Total}}
& \multicolumn{3}{c|}{\textbf{Accuracy (\%)}} 
& \multicolumn{3}{c|}{\textbf{Level of Detail (\%)}} \\
\cline{4-9}
& & & \textbf{M} & \textbf{P} & \textbf{N} 
& \textbf{M} & \textbf{P} & \textbf{N} \\
\hline
Zero-shot  & Claude-3.5 Sonnet & 22 & 5.3   & 47.4  & 47.4   & 15.8  & 47.4  & 36.8 \\
One-shot   & DeepSeek-V2.5 & 22 & 18.2  & 40.9  & \cellcolor{medGreen}\textbf{40.9}   & 31.8  & 50.0 & 18.2 \\
Few-shot   & Claude-3.5 Sonnet & 22 & 5.0   & 40.0  & 55.0  & 10.0  & 50.0 & 40.0 \\
GPA      & Claude Code       & 22 & 0.0   & 9.1   &\cellcolor{medOrange}\textbf{ 90.9}  & 0.0   & 18.2  &\cellcolor{medOrange} \textbf{81.8} \\
ArchView   & Claude-3.5 Sonnet & 22 & 18.2  & 40.9  & \cellcolor{medGreen}\textbf{40.9 }  & \textbf{50.0} & 36.4  & \cellcolor{medGreen}\textbf{13.6} \\
\hline
\end{tabular}
} 
\caption{Results for Human evaluation }
\label{tab:consensus_by_setting} 
\end{table}
AV showed the strongest alignment between automated and human evaluations. It achieved \textasciitilde59\% accuracy and \textasciitilde86.0\% level-of-detail success, outperforming OS and FS. Automated-human discrepancy reduced to 18.3 points, smaller than for FS, OS and GPA. As shown in Figure~\ref{fig:overall_comparison},  comparison of the approaches reveals that AV achieves superior performance in every metric other than Completeness.\\
\textbf{\textit{Statistical significance:}} we evaluated the generation quality by testing whether no significant difference exists across five approaches ($H_{0}$) which was weighed against the alternative that at least one approach would produce significantly different quality scores (the observed variations are statistically unlikely to have occurred by random chance, with $\alpha=0.05$)

The Friedman test revealed highly significant differences in generation quality across all 3Cs metrics: Clarity $(Q=196.44)$, Consistency $(Q=49.37)$, and Completeness $(Q=17.98)$. These results allow us to reject the null hypothesis. Post-hoc Wilcoxon tests demonstrated that AV consistently outperformed the other configurations in Clarity, showing a large effect size against the GPA $(r=0.829)$ and a medium effect over FS $(r=0.370)$. In Consistency, AV achieved significant gains with large effect sizes against the GPA $(r=0.596)$ and OS $(r=0.571)$. Notably, while the omnibus test for Completeness was significant, pairwise comparisons revealed that the differences were less pronounced; FS improved upon AV $(r=−0.421)$. This indicates that while AV excels in clarity and consistency, achieving completeness remains challenge.

\vspace{-13pt}
\begin{tcolorbox}[colback=blue!5!white, colframe=blue!95!white, colbacktitle=blue!95!white]
\small
    \textbf{Answer to RQ1}: ArchView consistently demonstrated best performance, followed by the Few-shot and One-shot approaches. In contrast, General Purpose agents exhibited the lowest efficacy.
\end{tcolorbox}
\vspace{-10pt}
\subsection*{RQ2: How do architectural notation and granularity affect view generation quality?}
\begin{figure}[h]
\centering
\includegraphics[width=0.8\columnwidth]{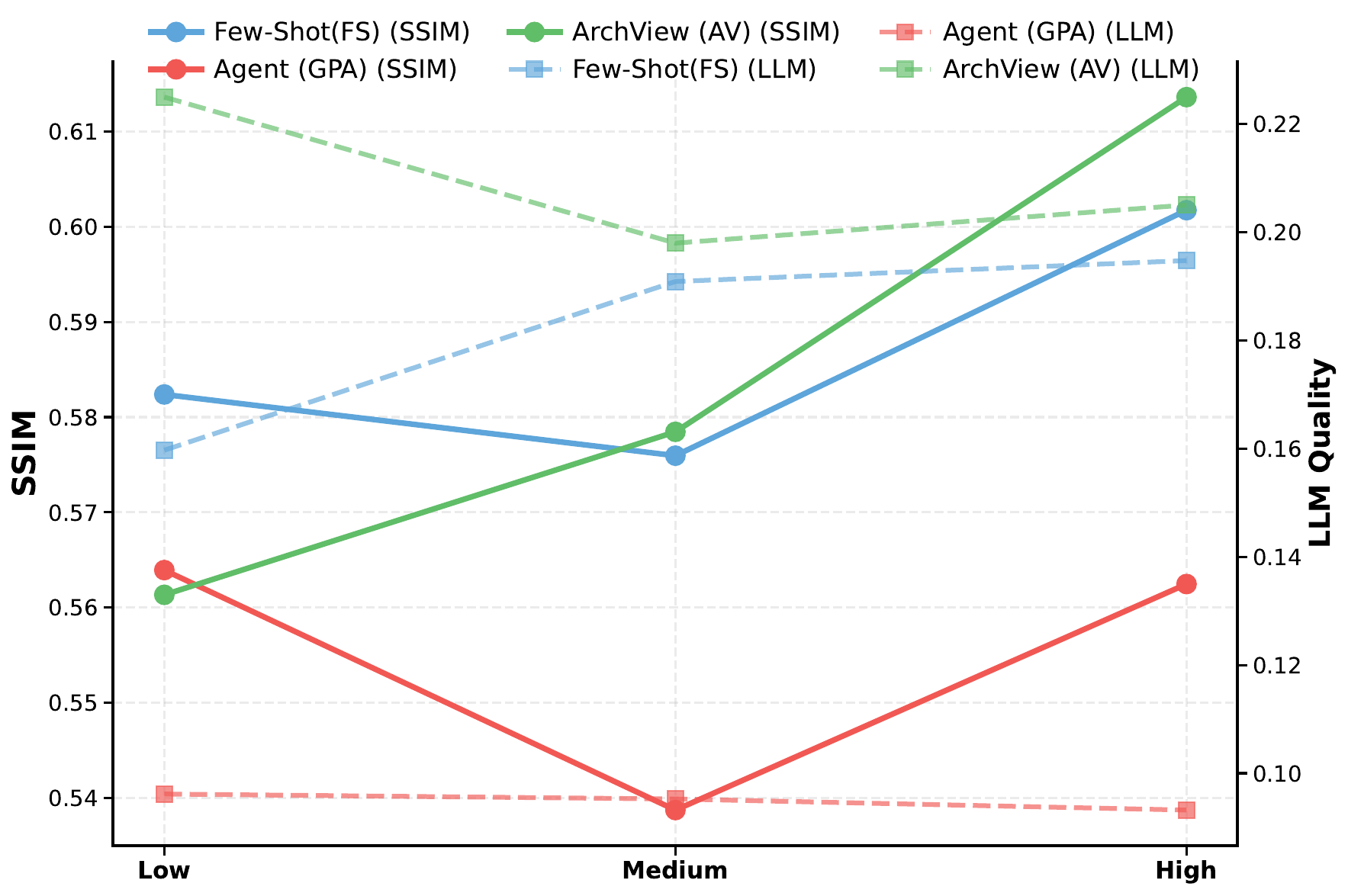}
\caption{Granularity trends for AV, FS, GPA. LLM Quality is the average of all LLM evaluated metrics}
\label{fig:granularity}
\end{figure}

This and the subsequent section focus on the FS, GPA, and AV approaches. We omit detailed analysis of the ZS and OS results here, as they yielded intermediate performance. Results reveals improvement of SSIM as notation complexity increased. Simple "boxes" notation achieved SSIM of only 0.534, while complex "boxes and arrows,icons and arrows" reached the highest SSIM 0.738 (+38\%). This suggests that richer notational vocabulary may provide LLMs with more structural cues for accurate view generation. \\
\begin{figure}[ht]
\centering
\includegraphics[width=9cm,height=5cm]{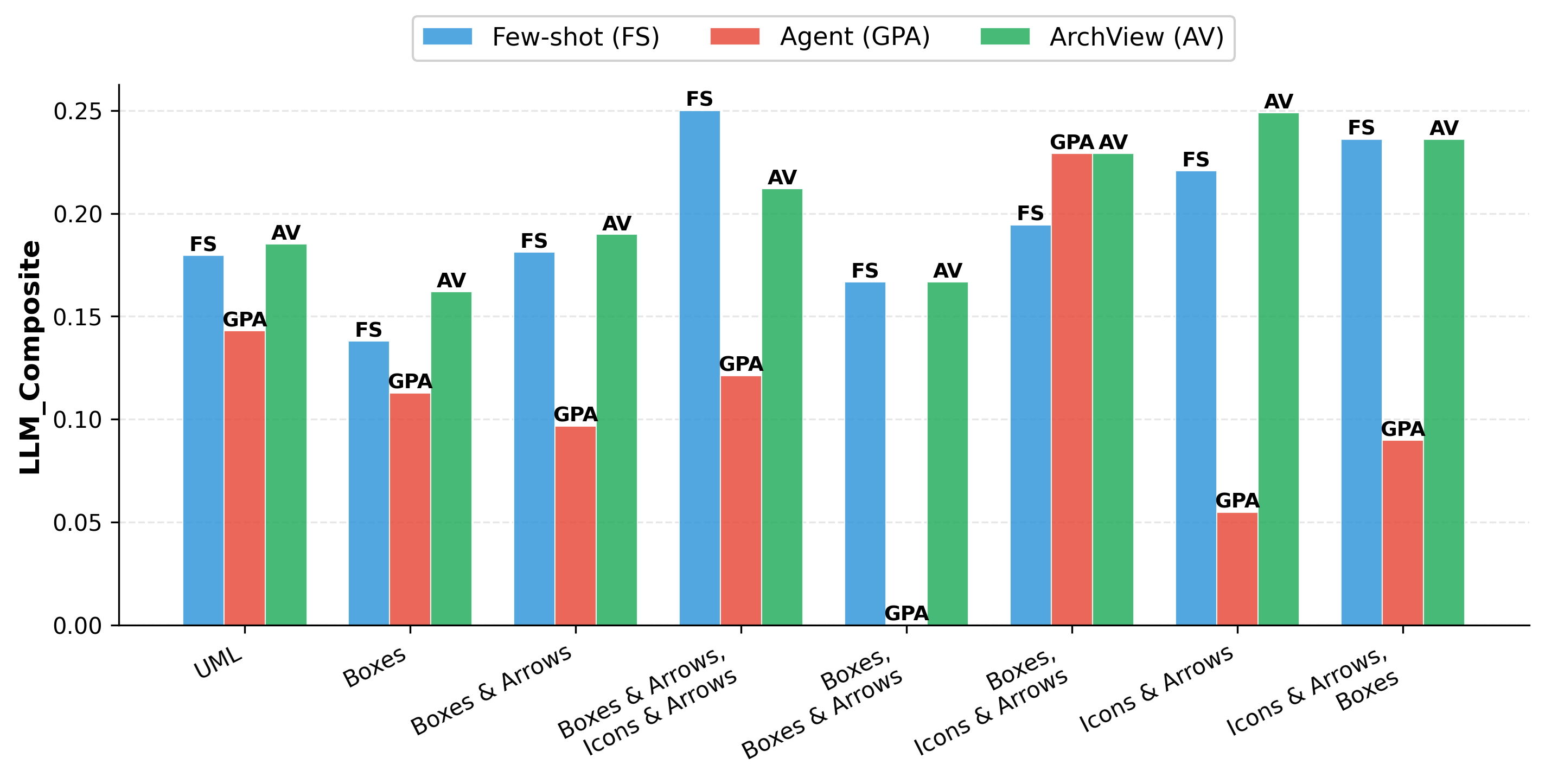}
\caption{LLM Quality across different notations}
\label{fig:notation}
\end{figure}
As shown in Fig.~\ref{fig:notation}, GPA's LLM quality (mean of all 3Cs rating) failure persisted across all notation types, suggesting that general coding agents are unable to leverage notational diversity regardless of visual complexity. Across the evaluated range of 0.138 to 0.250, AV and FS achieved their highest semantic performance using complex notations: ‘boxes and arrows, icons and arrows’ (FS: 0.250, AV: 0.212) and ‘icons and arrows’ (FS: 0.221, AV: 0.249).\\
Figure~\ref{fig:granularity} demonstrates distinct patterns across granularity levels. For high-granularity views, AV (SSIM 0.61, LLM quality 0.20) outperformed GPA and FS. SSIM improved from low to high granularity for AV and FS,
LLM quality scores declined as granularity increased, revealing a discrepancy between structural and semantic performance. GPA's LLM quality remained low across all granularities.

\textbf{\textit{Statistical significance:}} We formulated Architectural notation complexity has no significant effect  ($H_{0,\mathrm{N}}$) and notation complexity has a significant effect as the alternate hypothesis ($H_{1,\mathrm{N}}$). The Kruskal--Wallis test was highly significant for SSIM across all models leading to rejection of $H_{0,\mathrm{N}}$. Post-hoc analyses indicated that the \textit{icons\_and\_arrows}  significantly outperformed simple boxes, with large effect sizes.

In contrast, LLM Quality was not significantly affected by notation.
For Granularity, Granularity level (low, medium, high) has no significant effect on view generation quality was formulated as $H_{0,\mathrm{G}}$ while level of architectural granularity has a significant effect as alternate hypothesis $H_{1,\mathrm{G}}$. No statistically significant differences were observed for either SSIM or LLM Quality (all $p>0.05$), and thus $H_{0,\mathrm{G}}$ could not be rejected. It is worth noting that the distribution across granularity levels was uneven (Low: $n=34$, Medium: $n=140$, High: $n=199$), which may have reduced statistical power for detecting differences involving the smaller Low-granularity group. 
\vspace{-6pt}
\begin{tcolorbox}[colback=blue!5!white, colframe=blue!95!white, colbacktitle=blue!95!white]
\small
    \textbf{Answer to RQ2}: Richer notation improves view generation quality. In contrast, although observable differences in raw performance exist in the granularity comparison, these differences are not statistically significant.
\end{tcolorbox}
\vspace{-12pt}
\subsection{RQ3: How does performance vary across architectural concerns and quality attributes?}
\begin{figure*}[ht]
\centering
\includegraphics[width=\textwidth]{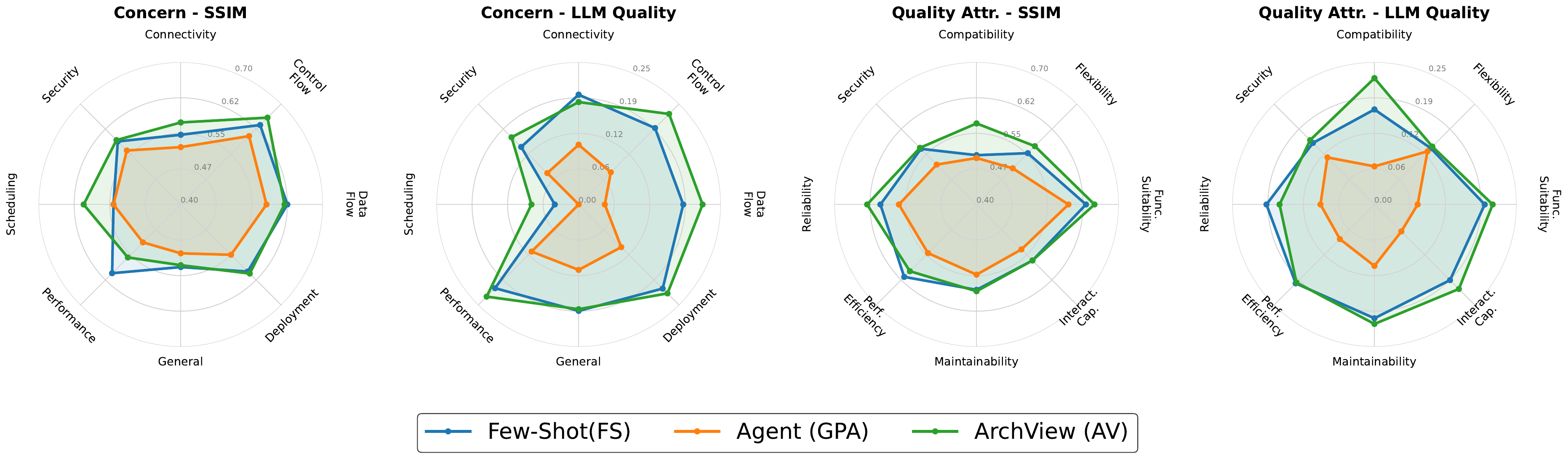}
\caption{Comparison of model performance across quality attributes and architectural concerns.}
\label{fig:combined_radar}
\end{figure*}

Figure~\ref{fig:combined_radar} presents the SSIM performance across architectural concerns. Control flow and Data flow achieved highest structural similarity. Performance views also benefited from AV (0.56) compared to GPA. Deployment showed moderate differentiation. General documentation proved most challenging view to be generated across all approaches. Although visually similar to other approaches in the SSIM radar, GPA showed severe semantic failure in the LLM quality ($<$\textasciitilde0.1) analysis.

Performance views achieved strongest semantic quality for AV, representing the best LLM quality. Control flow showed strong performance, as did data flow and deployment \textasciitilde0.21. Connectivity maintained moderate LLM quality. General documentation and security remained weaker. In Scheduling views, GPA's complete failure indicates fundamental challenges for temporal concerns. Control flow and data flow achieved balanced performance across both SSIM and LLM quality, making them most reliable for automated view generation. Performance views achieved highest semantic quality despite moderate SSIM, suggesting strong architectural understanding but visual presentation challenges. General documentation failed both visually and semantically.

As illustrated in Figure~\ref{fig:combined_radar}, SSIM performance varies across Quality Attributes (QA), with Functional Suitability achieving SSIM score (AV 0.649), followed by Reliability and Performance Efficiency, and consistent results for Maintainability (AV 0.583), whereas Flexibility, Security, and Interaction Capability demonstrate comparatively moderate performance.

From the QA's LLM quality radar (Figure~\ref{fig:combined_radar}), GPA showed failure across all quality attributes. Compatibility achieved highest semantic quality, despite lowest SSIM scores. Maintainability and interaction capability showed strong performance (AV 0.210). Functional suitability maintained solid quality. Performance efficiency and reliability show moderate.
The divergence between SSIM and LLM quality patterns reveals critical insights. GPA achieved moderate visual similarity (0.498-0.594) but worst semantic correctness (0.067-0.132), confirming pixel-level similarity does not guarantee architectural accuracy. Compatibility showed inverted pattern: lowest SSIM but highest LLM quality. Functional suitability showed balanced strength. Performance efficiency demonstrated moderate balance. Flexibility and security struggled on both dimensions, indicating generation challenges.\\
\textbf{\textit{Statistical Significance:}} We tested $H_{0,\mathrm{C}}$, which states that view generation performance is uniform across architectural concerns, against the alternative that some concerns achieve higher quality than others. Architectural concern type significantly affected SSIM but not semantic quality. Significant differences in structural similarity were observed for AV ($H=23.61$, $p<0.001$), FS ($p=0.002$), and GPA ($p=0.005$), leading to rejection of $H_{0,\mathrm{C}}$. Post-hoc analyses showed that Control Flow views significantly outperformed General Documentation across all models ($p \le 0.002$), with medium effect sizes ($\delta=0.33$--$0.39$).
Quality attribute type significantly influenced SSIM for GPA ($H=16.94$, $p=0.002$), ArchView ($H=15.79$, $p=0.003$), and Few-Shot ($H=13.55$, $p=0.009$), resulting in rejection of $H_{0,\mathrm{Q}}$ for visual similarity. A performance hierarchy emerged in which Functional Suitability and Security outperformed Flexibility, with the largest difference observed between Security and Flexibility for GPA ($\delta=+0.566$). In contrast, LLM Quality showed no statistically significant variation across QA.

\begin{tcolorbox}[colback=blue!5!white, colframe=blue!95!white, colbacktitle=blue!95!white]
\small
    \textbf{Answer to RQ3}: Control and Data Flow views demonstrated the high reliability. General documentation fails. For quality attributes, functional suitability achieves highest SSIM, while compatibility achieves highest LLM quality.
\end{tcolorbox}
\vspace{-9pt}
\section{Discussion}
\label{sec:discussions}

\subsection{Key Findings}

\begin{figure*}[htbp]
    \centering
    
    \begin{subfigure}{0.22\textwidth}
        \centering
        \includegraphics[width=\textwidth]{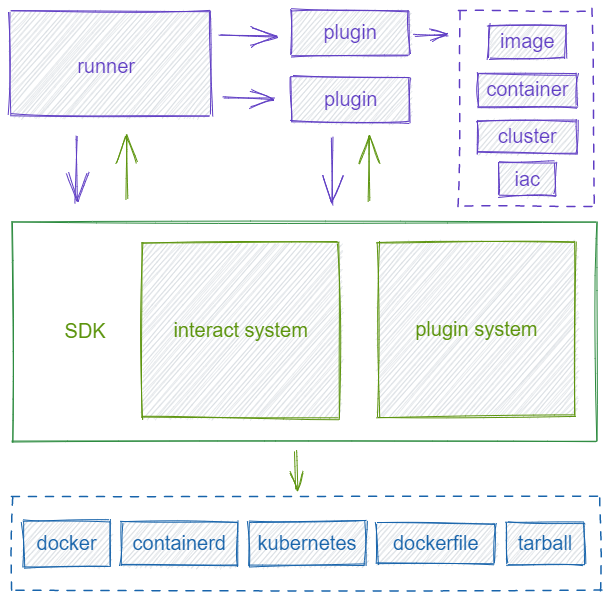}
        \caption{Ground Truth (Bad)}
        \label{fig:sub1}
    \end{subfigure}
    \hfill 
    \begin{subfigure}{0.28\textwidth}
        \centering
        \includegraphics[width=\textwidth]{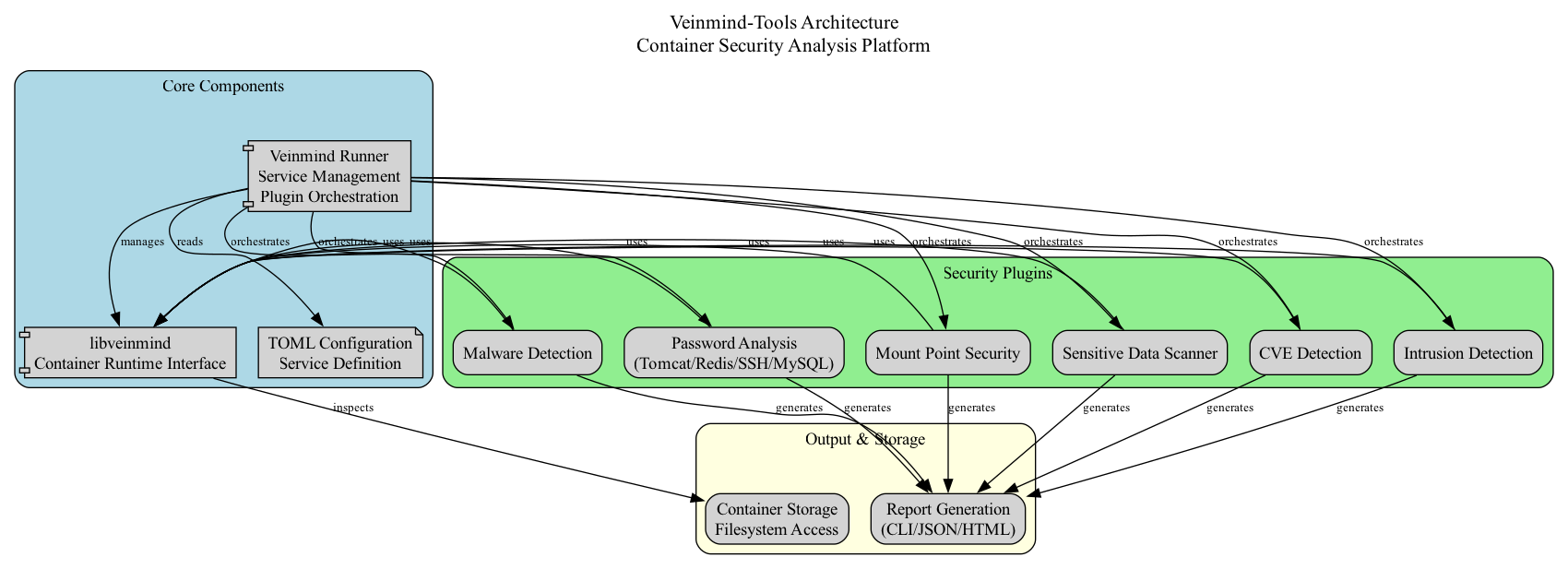}
        \caption{ArchView (Bad)}
        \label{fig:sub2example}
    \end{subfigure}
    \hfill
    \begin{subfigure}{0.22\textwidth}
        \centering
        \includegraphics[width=\textwidth]{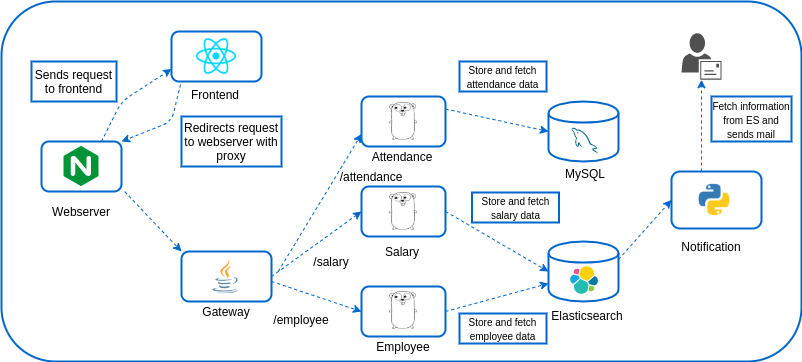}
        \caption{Ground Truth (Good)}
        \label{fig:sub3}
    \end{subfigure}
    \hfill
    \begin{subfigure}{0.22\textwidth}
        \centering
        \includegraphics[width=\textwidth]{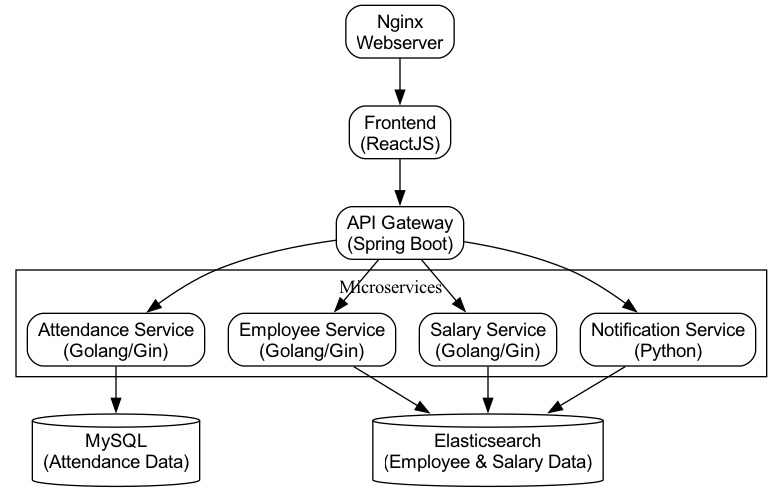}
        \caption{ArchView (Good)}
        \label{fig:sub4}
    \end{subfigure}
    
    \caption{Examples of Ground Truth and ArchView Views}
    \label{fig:example}
\end{figure*}

\textbf{LLMs generate syntactically valid but semantically flawed architecture views.} Although literature suggests that generating views purely from source code is inherently difficult because architectural intent is rarely explicitly encoded~\cite{fairbanks2023software}, recent advancements have begun to surmount this challenge. Our evaluation across 340 repositories revealed that all three LLMs successfully produced architecture views, with error rates remaining below 11\%. Our results indicate that structural correctness does not necessarily imply architectural accuracy or appropriate granularity. While 266 of the 340 generated views closely matched the ground truth in terms of clarity (AV), they struggled significantly with completeness and consistency. This suggests that while generated views can mimic the structure of ground truth diagrams, they often fail to comprehensively capture the underlying architectural information from source code (Fig. ~\ref{fig:example}).\\
\textbf{Specialized domain-specific agents are essential; neither LLMs nor general purpose agents suffice for architectural tasks.} Our experiments revealed a clear performance hierarchy: while LLMs plateaued at 31\% clarity failures (FS), GPA failed at 71.8\%, custom agent(AV) reduced the failure rate to 22.6\%. Claude Code generated code-level granularity with excessive components, poinitng to an inability of architectural abstraction. Custom-built agent(AV) achieved 78\% clarity success, 50.0\% level-of-detail success. Automated-human evaluations alignment improved substantially (18.3pp gap vs. 24-54pp). These results confirm that architectural understanding requires the integration of domain-specific knowledge, concern specifications, and abstraction.\\
\textbf{View Generation quality varies across view types.} We found that generation quality improves with notational complexity; richer visual vocabularies (e.g., icons and arrows) yielded a 38–67\% SSIM improvement over simple box-and-line notations. Custom-built agent (AV) demonstrated superior capability in handling all abstractions, particularly for Control Flow and Data Flow concerns. Notably, certain quality attributes like Compatibility achieved high semantic scores despite low visual fidelity, indicating that while the model captures the architectural logic, it struggles to spatially organize complex dependency graphs. Future works can on deeper understanding on the failure of specific type of views and developing approaches which can capture them. 

\subsection{Implications for Researchers}
\textbf{Source code summarization remains a critical bottleneck.} While existing literature has successfully addressed the mapping of source code to low-level representations~\cite{babaalla2025llm}, our results demonstrate that generating high-level architectural views based on source code summarization remains non-trivial. We found that view generation quality is strictly bound by summarization accuracy; when abstraction fails, views diverge regardless of the model used (Fig. ~\ref{fig:sub2example}). This dependence creates a significant bottleneck. Future research must advance in code base summarization to extract high-level reasoning.\\
\textbf{Evaluation metrics for architecture views require advancement.} While Camara et al. highlight the scarcity of benchmarks for high-level modeling~\cite{Cámara2024}, and Bouali et al. demonstrate that LLMs can reliably grade low-level UML artifacts~\cite{Bouali}, our results indicate that existing evaluation techniques remain insufficient for generated architectural views. We observed a critical divergence between structural and semantic metrics. For instance, while ArchView’s SSIM (0.7) correlated with quality, the Agent approach achieved a misleading SSIM of 0.6 despite suffering from near-total semantic failure (0.09 LLM quality). Furthermore, unlike the high human-correlation reported in educational grading tasks, our LLM-as-a-Judge demonstrated to be too strict occasionally(26\%), frequently rejecting valid architectural abstractions. Consequently, evaluation frameworks must move beyond pixel-level or rigid textual comparisons to include architecture-specific metrics for abstraction appropriateness, stakeholder comprehensibility, and concern coverage.\\
\textbf{The formal-semiformal notation gap limits practical applicability.} LLMs consistently generated formal PlantUML diagrams with strict notational standards, while real-world practices frequently employ semi-formal or informal notations with flexible visual conventions~\cite{malavolta2012industry} (Figs.~\ref{fig:sub1} and~\ref{fig:sub3}). Eliminating the constraint of strictly formal notation likely contributes to ArchView's consistent success across diverse notational styles. Notation complexity analysis revealed performance improved with richer notation: simple "boxes" achieved SSIM 0.54 while complex "boxes and arrows, icons and arrows" reached SSIM 0.74 (Figure~\ref{fig:notation}). This may also be due to the lack of datasets with formal notations (at higher levels of abstractions - 15/340) calling for more dataset collection.
\vspace{-7pt}
\subsection{Implications for Practitioners}

\textbf{Leverage concern and quality attribute insights for targeted view generation.} While literature establishes that diverse architectural views are indispensable for reasoning about quality-related variability and abstractions which are not explicitly encoded in source code~\cite{fairbanks2023software, galster2013variability}, our study provides a preliminary mapping of the view types appear most amenable to automated generation. Our results suggest that standard structural views, particularly Control Flow addressing Functional Suitability, proved highly reliable, whereas concerns requiring deeper reasoning about variability such as Deployment and Performance necessitated the Custom-built (AV) for best results. Based on these performance disparities, we provide an observations analysis for practitioners to target amenable concern-quality attribute combinations\footnote{Observations table: \url{https://zenodo.org/records/18772573}}. However, the persistent challenge in generating "General" concern indicates that holistic synthesis remains a challenge, suggesting that future work could benefit from focusing on the semantic integration of broader architectural contexts.
\textbf{Adopting LLM-Based View Generation.} Literature indicates that rapid software evolution, frequent changes, and the lack of developers awareness are the most common causes occurred in architecture degradation~\cite{architecturedegradation}. Despite identified limitations, our results demonstrate LLM-based architecture view generation has reached practical viability for specific use cases. With 78\% clarity success and 50\% level-of-detail success for custom-built agents, and even few-shot prompting achieving 70\% clarity success, these tools can meaningfully accelerate architectural documentation workflows when applied with appropriate caution and oversight.

\textbf{Generated views aid architectural knowledge extraction despite quality limitations.} Even imperfect LLM-generated views has the potential to accelerate comprehension for unfamiliar or legacy systems. They serve as effective starting points for system understanding, dependency mapping, and gap detection. Practitioners should leverage these outputs to bootstrap documentation and review workflows, prioritizing rapid insight over immediate perfection.

\section{Threats to validity}
\label{sec:threatstovalidity}

We follow the categorization provided by Wohlin et al. \cite{wohlin2012experimentation} to identify potential risks to our study and detail the efforts taken to mitigate them.

\textbf{External validity:} Our selection of LLMs and agents, which may not be fully representative of AI landscape. We partially mitigated this by systematically selecting models from the lmarena leaderboard. Our evaluation of GPA was limited to Claude Code. This poses a threat to generalizability across all coding agents. The potential for data leakage remains a partially mitigated threat; since LLM training data is opaque, we cannot definitively rule out that our open-source evaluation repositories \cite{schaeffer2023pretraining} were included in their training sets. We partially mitigate this as 113 repositories (33.43\%) in our dataset were created or significantly updated after 2023, which post-dates the known training cutoffs for the models used in our study.

\textbf{Internal validity:} To address configuration bias, we mitigated potential inconsistencies by using default configurations for all models and maintaining identical parameters across every experimental run. A primary threat in our study is the inherent reliability of LLM-as-a-Judge. Relying on an LLM to evaluate complex architectural views is a significant threat, as the model’s internal evaluation logic may deviate from the nuanced reasoning required for software engineering documentation. We partially mitigated this by providing a highly structured rubric. Closely related is the threat of self-preference bias \cite{Wataoka2024SelfPreferenceBI}, where the evaluator model might favor its own outputs or specific formatting styles. We partially mitigated this by complementing automated scores with a blinded human evaluation. Furthermore, the choice of diagramming notation posed a risk to output quality; our pilot study revealed that Mermaid lacked robust component diagram support and produced higher error rates. We mitigated this by standardizing PlantUML as our primary tool. Due to the small sample size of the human evaluation, statistical tests were not performed. 

\textbf{Construct validity:} We relied on manually coded architecture views from Migliorini et al. \cite{10.1007/978-3-031-70797-1_27} as our ground truth. We mitigate the threat of not having a clear ground truth by using this specific dataset. Regarding the decision of our qualitative assessment, the human evaluation phase operated with an 85\% confidence level and a $\pm 15\%$ margin of error. While this margin limits our ability to detect minute performance variances, we partially mitigated the risk of subjective error by employing a blinded design.

\vspace{-2mm}
\section{Related Work}
\label{sec:related}

The extraction of software architecture from source code has been a long-standing research theme~\cite{alshuqayran2016, geiger2018,suggestingsoftwarearchitecture}, Zdun et al.~\cite{litreviewontraceability}  highlight architectural views importance to derive abstractions from source code. Building on Fairbanks conceptualization of software architecture ~\cite{fairbanks2023software}, view consistency is essential for preserving architectural integrity, especially as systems evolve, degrade, or accumulate technical debt~\cite{babaalla2025llm}. 

Recent advances in Generative AI have motivated several efforts exploring the role of LLMs in software architecture and software engineering~\cite{hou2024large,smellextraction,navigatingcomplexity,futuresoftware}. Survey papers and industry studies report the adoption of LLMs for activities such as architectural decisions, code generation, and architecture knowledge management~\cite{jahic2024state,surveyonLLMs}. Jahić et al. empirically investigated LLM adoption in software engineering; their work uses C4 and UML to represent LLM-generated designs for systematic comparison with human designs~\cite{jahic2024state}. Other works have explored using LLMs to generate architectural components (FaaS)~\cite{architecturalcomponents}. Complementary to these trends, prior work has also investigated automated generation of Architecture Decision Records (ADRs)~\cite{architecturaldesigndecisions}. Recent studies suggest LLMs can improve architectural knowledge management through ADR generation and query-based access~\cite{akmmanagement}. 

Several studies have focused on using LLMs to generate low-level models, particularly UML artifacts. Cámara et al. proposed one of the first conceptual benchmarking frameworks for evaluating LLMs in software modeling using class diagrams~\cite{Cámara2024}. De Bari et al. investigated the ability of LLMs to generate class diagrams from textual requirements~\cite{de2024evaluating}. Bouli et al. explored using LLMs to automatically grade UML class diagrams by converting student submissions into textual descriptions~\cite{Bouali}. Conrardy et al. investigated the use of LLMs to convert hand-drawn UML class diagram images into formal, machine-readable models~\cite{conrardy2024image}. Ferrari et al. examined how LLMs can support incremental and interactive diagram definition during requirements engineering~\cite{ferrari2024model}. Babaalla et al. proposed an MDA-aligned pipeline that transforms textual specifications into UML models and code~\cite{babaalla2025llm}. Finally, the use of generative AI for architecture knowledge management is an emerging area, with growing interest in applying LLMs to support architectural documentation, rationale capture. 

Departing from these studies, which primarily focus on deriving low-level representations, our work investigates whether LLMs and agentic workflows can generate high-level architectural abstractions. To the best of our knowledge, this is the first study to systematically evaluate LLM-based approaches for architecture view generation. 
\section{Conclusion and future work}
\label{sec:conclusionandfutureworks}
This study evaluates the capabilities of LLM-based approaches on automated generation of architecture views from source code. Using 340 open-source repositories across 13 experimental configurations, we assessed 4,137 generated views through automated metrics, LLM-as-a-Judge evaluation, image similarity analysis, and human expert assessment.

Future work should address several directions. First, advancing repository summarization techniques. Second, implementing Retrieval-Augmented Generation (RAG) to better incorporate external ground-truth knowledge and enhance repository-wide contextualization. Third, exploring multi-agent collaboration and architecture-specific evaluation metrics. 

\section{Data Availability}
All experimental data, scripts, prompt templates, and results are available in our replication package: \url{https://zenodo.org/records/18772573}
\section*{Acknowledgment}
This material is partly supported by the Google Cloud Research Credits program with the EDU credit Grant 417805117.

\bibliographystyle{ieeetr}
\bibliography{references}
\end{document}